# Bremsstrahlung radiation from the interaction of short laser pulses with dielectrics

G. M. Petrov, J. P. Palastro, and J. Peñano
*Naval Research Laboratory, Washington DC 20375-5346, USA*

**Abstract**

An intense, short laser pulse incident on a transparent dielectric can excite electrons from valence to the conduction band. As these electrons undergo scattering, both from phonons and ions, they emit bremsstrahlung radiation. Here we present a theory of bremsstrahlung emission appropriate for laser pulse-dielectric interactions. Simulations of the interaction, incorporating this theory, illustrate characteristics of the radiation (power, energy and spectra) for arbitrary ratios of electron collision frequency to radiation frequency. The conversion efficiency of laser pulse energy into bremsstrahlung radiation depends strongly on both the intensity and duration of the pulse, saturating at values of about $10^{-5}$. Depending on whether the intensity is above or below the damage threshold of the material, the emission can originate either from the surface or the bulk of the dielectric respectively. The bremsstrahlung emission may provide a broadband light source for diagnostics.

## 1. Introduction

The wide availability of high intensity, ultrashort pulse lasers has led to an explosion of research in ultrafast laser-material interactions. The irradiation of transparent, solid dielectrics, in particular, has been investigated in several contexts, including non-thermal heating and laser ablation, [1,2,3,4], high precision hole drilling [5], and nano-cavity formation [6,7]. Regardless of context, the interaction evolves along a common sequence of processes: (i) production of conduction-band electrons via optical field and collisional ionization; (ii) deposition of laser energy in conduction band electrons; and (iii) transfer of energy from conduction band electrons to the lattice leading to melting, surface ablation or micro-explosions within the dielectric.

These processes are not, however, exhaustive. For instance, radiative phenomena, such as supercontinuum and high harmonic generation, regularly occur during the interaction [8,9,10,11,12]. Of interest here is collisional radiation, or bremsstrahlung. While bremsstrahlung radiation is commonly associated with plasmas, the underlying mechanism, Larmor radiation from the acceleration of an electron during a collision, occurs in laser-heated dielectrics as well. In plasma, free electrons undergo collisions with ions; in dielectrics, conduction band electrons (CBEs) scatter from ions or the lattice (electron-phonon scattering).

Bremsstrahlung radiation from plasmas has been studied extensively, often providing an indispensable diagnostic tool. Thermal bremsstrahlung radiation has been used, for example, to diagnosis hot electron populations in inertial confinement fusion experiments [13,14], provide information about star formation in starburst galaxies [15,16], and to extract radiation temperatures in discharges [17]. In all of these cases, the plasma can be considered both fully ionized and weakly collisional, i.e. the collision frequency, $\nu$, is much smaller than the radiation frequency, $\omega$. These conditions greatly simplify calculations of the bremsstrahlung spectrum and emitted power, permitting the use of well-known formulas for the spectral intensity [18] or power emitted per unit volume [15].

In contrast, bremsstrahlung radiation from CBEs in dielectrics has received little attention [19]. Transient ionization and a range of collision frequencies typify the laser pulse-dielectric interaction. This invalidates the simplifications of full ionization and weak collisionality used in plasmas, such that the well-known formulas are not applicable. As we will show, the bremsstrahlung emission exhibits a broadband spectrum, spanning microwave to ultraviolet frequencies, with the collision frequency often falling in between. Thus a theory of bremsstrahlung radiation appropriate for dielectrics must treat arbitrary ratios of collision frequency to radiation frequency, $\nu/\omega$.

Furthermore, the theory must include both electron-phonon and electron-ion collisions. To our knowledge, such a theory has yet to be presented.

The purpose of this manuscript is to develop a theory and numerical model appropriate for the bremsstrahlung emission from short laser pulse irradiated dielectrics, and to illustrate the characteristics of bremsstrahlung radiation at conditions relevant to laser pulse-dielectric interactions, i.e. radiation emitted by cold (a few eV), dense (solid density) electrons in the conduction band that undergo both phonon and ion collisions. Characterizing laser-generated bremsstrahlung radiation from dielectrics may motivate its potential use as a diagnostic.

The remainder of this manuscript is organized as follows. Section 2 provides a brief overview of the emission characteristics. In section 3, we develop the general theory of bremsstrahlung radiation for arbitrary ratio of collision frequency to radiation frequency. The frequency and temperature scalings of a representative radiation spectrum are discussed therein. Section 4 details an implementation of the bremsstrahlung theory for simulations. Section 5 presents our laser pulse-dielectric interaction simulations that incorporate the bremsstrahlung theory. For a representative dielectric, we use $SiO_2$, which has well-established properties and electron collision data, and has been widely used for numerical simulations [1,3,6,20,21,22,23]. The simulations results include the bremsstrahlung spectra, power, energy, and conversion efficiency. For laser pulse intensities below and above the damage threshold, we find the dielectric is predominately a *volumetric* or *surface* emitter, respectively. At the end of the section, we present scalings of the conversion efficiency as a function of pulse intensity for a short (fs) and long (ps) pulse. Section 6 summarizes our findings.

## 2. Emission Characteristics

Prior to presenting the theory and simulations of bremsstrahlung emission, it is instructive to discuss qualitative features based on what is known about short pulse-dielectric interactions [22,23]. As an example, we consider a pulse with a wavelength $\lambda_0 = 800\ \text{nm}$, full width at half maximum (FWHM) duration $T_{FWHM} = 1\ \text{ps}$, and an intensity, $I_L$, near the breakdown threshold, $I_{thr} \sim 2\times10^{17}\ \text{W}\cdot\text{m}^{-2}$. When incident on the dielectric, the pulse excites a thin surface layer of electrons to the conduction band. The CBEs have a density, $n_e$, near the critical density, $n_{cr} = \pi / r_e\lambda_0^2 = 1.7\times10^{27}\ \text{m}^{-3}$ where $r_e$ is the classical electron radius, and a temperature $k_B T_e \sim 5\ \text{eV}$, where $k_B$ is Boltzmann's constant. The layer thickness is about one skin depth, $\ell_{skin} = (4\pi n_e r_e)^{-1/2} \sim 100\ \text{nm}$.

As we show in the next section, the power emitted per unit volume into bremsstrahlung radiation is given, in the collisional limit, by $P_{brems} \cong (4\alpha / \pi\hbar^2 mc^2)(k_B T)^4 n_e / \nu(T_e)$, where $\alpha$ is the fine structure constant, $\hbar$ the Planck constant, $m$ the electron mass, and $c$ the speed of light. The total collision frequency, $\nu$ , includes both electron-phonon and electron-ion processes, but is dominated, for nearly all parameters of interest, by electron-phonon scattering. The conversion efficiency of pulse energy into bremsstrahlung radiation, $\eta_{brems} \cong P_{brems} \ell_{skin} / I_{thr}$ , can then be estimated by setting $n_e = n_{cr}$ and $\nu = 3\times10^{15}\ \mathrm{s}^{-1}$ [20], providing $\eta_{brems} \sim 10^{-6}$. The modest efficiency results predominately from the low electron temperature.

The bremsstrahlung power spectrum is broadband, spanning microwave to ultraviolet frequencies, and peaks near the cut-off frequency, $\omega_{\max} \sim k_B T_e / \hbar$ or $\lambda_{\max} \sim 250\,\mathrm{nm}$, close to the visible range. The rapid recombination of CBEs, e.g. 150 fs in $SiO_2$, provides an upper limit to the duration of emissions for very short laser pulses. The bremsstrahlung emission would therefore appear as an ultrafast flash of broadband light.

### 3. Bremsstrahlung Radiation Theory

We now develop the theory of bremsstrahlung radiation emitted by thermal electrons in the conduction band of a dielectric. For simplicity, we consider a dielectric that maintains its lattice structure throughout the interaction, neither melting nor ablating. This assumption is made in order to avoid complications associated with phase transitions and allow for a well-defined electron-phonon collision frequency. In general, the electrons are not in thermodynamic equilibrium with the dielectric lattice, having $T_e >> T_\ell$, where $T_\ell$ is the lattice temperature. For an overview of the mechanisms for excitation to the conduction band and collisional processes refer to Refs [1].

Conventional treatments of Bremsstrahlung radiation consider the weakly collisional limit, in which the collision frequency, $\nu$, is assumed much smaller than the radiation frequency, $\omega$ : $\nu / \omega << 1$. For short pulse heated dielectrics, the radiation frequency often falls in the opposite limit. To accommodate a broad range of emitted frequencies, we treat the more general case in which the ratio of collision frequency to radiation frequency is arbitrary. Analytical expressions for the emission, absorption coefficient, optical depth, and radiation intensity will be reduced to the two limiting cases: weak $(\nu / \omega << 1)$ and strong $(\nu / \omega >> 1)$ collisionality. For convenience, the formulas and notation are written in SI units and follow closely Refs. [15,18].

**(a) Radiative transfer**

The monochromatic radiation intensity, the energy emitted per unit area, time, frequency interval $d\omega$ and solid angle $d\Omega$, evolves according to the radiation transport equation

$$\frac{dI_\omega}{ds} = j_\omega - a_\omega I_\omega, \tag{1}$$

where $s$ represents a possible path for the radiation, $j_\omega$ the emission coefficient, and $a_\omega$ the absorption length. The emission coefficient, $j_\omega$, characterizes the radiation source—in our case the bremsstrahlung emission of CBEs within an infinitesimal volume element of the dielectric. Calculation of the emission coefficient requires averaging the spontaneous emission rate for bremsstrahlung, $\eta_\omega(\varepsilon)$, over the electron energy distribution, $f(\varepsilon)$, and multiplying by the electron density:

$$j_\omega = n_e \int_0^\infty f(\varepsilon)\eta_\omega(\varepsilon)d\varepsilon\,. \tag{2}$$

The spontaneous emission rate, the energy radiated by a single electron with kinetic energy $\varepsilon$ per unit time, frequency interval, and solid angle is given by [18]

$$\eta_\omega(\varepsilon) = \frac{\alpha}{3\pi^2}\frac{\varepsilon}{mc^2}\mathrm{h}\nu(\varepsilon)\frac{1}{1+\nu^2(\varepsilon)/\omega^2}\,. \tag{3}$$

Note that $\nu(\varepsilon)$ and $f(\varepsilon)$, and hence $j_\omega$ depend implicitly on both space and time through the electron density and temperature. For the remainder, we assume a Maxwellian kinetic energy distribution: $f(\varepsilon) = N^{-1}\varepsilon^{1/2}\exp(-\varepsilon/k_B T_e)$, where the normalization $N = \frac{1}{2}\pi^{1/2}(k_B T_e)^{3/2}$ ensures $\int_0^\infty f(\varepsilon)d\varepsilon = 1$ and the average energy $\bar{\varepsilon} = \int_0^\infty \varepsilon f(\varepsilon)d\varepsilon = \frac{3}{2}k_B T_e$.

As an inverse process to emission, the CBEs can also absorb radiation. This is captured in Eq. (1) by the absorption length:

$$a_\omega = \frac{j_\omega}{B_\omega(T)}\ , \tag{4}$$

where $B_\omega(T) = (\hbar\omega^3/8\pi^3c^2)(e^{\hbar\omega/k_B T}-1)^{-1}$ is the black body radiation spectrum for a temperature *T*. Equation (4) can be derived by requiring energy conservation in the limiting case of a radiator in thermodynamic equilibrium with its environment: $I_\omega = B_\omega(T)$. To ensure energy conservation in this limit, the absorption must balance emission, such that $j_\omega - a_\omega I_\omega = 0$ or $a_\omega B_\omega(T) = j_\omega$ (see Eq. (1)).

With the absorption coefficient, one can define the optical depth, $d\tau_\omega$, over the path length

$ds$, as

$$d\tau_\omega = a_\omega ds. \tag{5}$$

The optical depth plays a central role in determining the transport of radiation through and out of the CBE region. To illustrate this, we consider Eq. (1) for the simple example of a flat, uniform radiator of length $L$ (infinite in the other dimensions). This is an excellent approximation of the CBE layer created at the dielectric surface for pulse intensities near or above the breakdown threshold (the CBE region has a thickness of $L \sim \ell_{skin} \sim 100\,\text{nm}$ and a width comparable to the pulse spot size $w \sim 100\,\mu\text{m}$ ). Integrating Eq. (1) for the flat, uniform radiator, we find

$$I_\omega = \frac{j_\omega}{a_\omega}(1 - e^{-\tau_\omega}). \tag{6}$$

Equation (6) has two limiting cases. An optically "thin" radiator satisfies $\tau_\omega << 1$, such that

$$I_\omega \cong j_\omega L. \tag{7}$$

In this case, the radiation leaving the CBE region is simply the radiation produced inside that region. An optically "thick" radiator, on the other hand, satisfies $\tau_\omega >> 1$, such that

$$I_\omega \cong \frac{j_\omega L}{\tau_\omega}. \tag{8}$$

Here the radiation is strongly absorbed, reducing the intensity leaving the CBE region by a factor $\tau_\omega = a_\omega L$, the optical depth of the media. Note that the frequency dependence of $\tau_\omega$ implies a radiator can be optically thin at one frequency and thick at another.

While this example illustrates the important limits, the physical thickness of the CBE region can vary widely: from a thin 100 nm layer just below the surface of the dielectric to a 1 mm swath trailing the laser pulse, depending on whether the pulse parameters are above or below the breakdown threshold respectively. A more general treatment for the emission from physically thick CBE regions is included in our simulations and will be discussed in the next section.

Through the spontaneous emission rate, the optical depth and emission coefficient for a particular radiator will depend on the properties of the collision frequency. Two properties distinguish the collision frequency of CBEs in dielectrics to that of traditional plasmas. First, the dominant collisions are electron-phonon, which depend solely on the electron temperature; the density-dependent Coulomb collisions provide only a small contribution. Second, numerical simulations will show that the peak electron temperature, ~a few eV, is relatively insensitive to laser pulse parameters. This insensitivity and the weak density dependence imply that the combined CBE collision frequency (electron-phonon + electron-ion) is nearly constant. As a result, delineation of the weakly ($\nu << \omega$) and strongly ($\nu >> \omega$) collisional regimes is determined predominately by the radiation frequency, and not the parameters of the interaction. For a rough demarcation in the following, we can use a typical electron-phonon collision frequency, $\nu \cong 3\times10^{15}\ s^{-1}$ [20].

**(b) Weak collisionality**

For short pulse heated dielectrics, the weakly collisional limit, $\nu << \omega$, applies to ultraviolet (UV) or shorter wavelength radiation. In this limit, the spontaneous emission rate becomes independent of frequency: $\eta_\omega(\varepsilon) = (\hbar\alpha / 3\pi^2 m_e c^2)\varepsilon\nu(\varepsilon)$. The complex energy dependence of $\nu(\varepsilon)$ precludes a direct analytical integration of Eq. (2) for $j_\omega$. We can, however, estimate the integral by evaluating $\eta_\omega$ at the average energy, providing $j_\omega \cong (\hbar\alpha / 3\pi^2 m_e c^2) n_e \bar{\varepsilon}\nu(\bar{\varepsilon})$. A more rigorous treatment, taking into account electrons with $\varepsilon \neq \bar{\varepsilon}$ by assuming their contribution to the frequency spectrum is proportional to their number, yields an additional factor $j_\omega \sim e^{-\hbar\omega/k_B T_e}$. For a Maxwellian kinetic energy distribution, we then have:

$$j_\omega \cong \frac{\alpha}{2\pi^2}\frac{k_B T_e n_e}{m_e c^2}\hbar\nu(T_e)e^{-\hbar\omega/k_B T_e} \tag{9}$$

$$a_\omega \cong \frac{\omega_p^2}{\omega^2}\frac{\nu(T_e)}{c}. \tag{10}$$

The absorption length recovers the inverse-bremsstrahlung absorption length commonly used in radiation-tenuous plasma interactions. The optical depth resulting from Eq. (10), $\tau_\omega = a_\omega L$, increases rapidly with decreasing frequency, $\tau_\omega \sim \omega^{-2}$. This implies that low frequencies (but still larger than $\nu$) are preferentially absorbed. For laser pulse heated dielectrics, this preferential absorption would be significant in a very limited frequency range. The range, $\nu << \omega \leq k_B T_e / \hbar$, is bounded below by the relatively large collision frequency, $\nu \sim 3\times10^{15}$ s$^{-1}$, and above by the exponential fall off of the emission, $k_B T_e / \hbar \sim 6\times10^{15}$ s$^{-1}$. By integrating $j_\omega$ over all frequencies and solid angle, we obtain the total power of bremsstrahlung radiation emitted per unit volume: $P_{brems} \cong (2\alpha / \pi m_e c^2)(k_B T_e)^2 n_e \nu(T_e)$. Note that this formula quantifies emission *within* the CBE region, while the frequency integral of $I_\omega$, which we return to below, quantifies the radiation *escaping* the CBE region. The expression for $P_{brems}$ reproduces the traditional plasma result when the Spitzer electron-ion collision frequency is substituted for $\nu(T_e)$.

**(c) Strong collisionality**

The strongly collisional limit, $\nu >> \omega$, applies for radiation with wavelengths approximately satisfying $\lambda > 1\,\mu$m. Following the procedure from subsection (b) above, the spontaneous emission rate reduces, in the limit $\nu >> \omega$, to $\eta_\omega(\varepsilon) \cong (\hbar\alpha / 3\pi^2 m_e c^2)\omega^2\varepsilon / \nu(\varepsilon)$. For $j_\omega$ and $a_\omega$ we then find:

$$j_\omega \cong \frac{\alpha}{2\pi^2}\frac{k_B T_e n_e}{m_e c^2}\frac{\hbar\omega^2}{\nu(T_e)}e^{-\hbar\omega/k_B T_e} \tag{11}$$

$$a_\omega \cong \frac{\omega_p^2}{c\nu(T_e)}. \tag{12}$$

In contrast to the weakly collisional limit, the absorption length, and hence the optical depth is independent of frequency. Integration of (11) yields the total power of bremsstrahlung radiation emitted per unit volume: $P_{brems} \cong (4\alpha / \pi\hbar^2 mc^2)(k_B T)^4 n_e / \nu(T_e)$. The power scales as $P_{brems} \sim T_e^4$ reminiscent of black body radiation with a temperature $T_e$, suggesting that strong collisionality results in an optically thick radiator. A summary of the formulas in the two limiting cases is provided in Table 1.

**(d) Radiation intensity spectrum and absorption coefficient**

To provide insight into the emission and absorption properties of the CBEs, we continue by examining qualitative features of the bremsstrahlung spectrum. For this purpose, it is sufficient to consider two quantities, which play central role in radiation emission and absorption: the radiation intensity $I_\omega$ and absorption coefficient $a_\omega$. The exact frequency scaling of $I_\omega$ depends on the optical depth of the radiator. However, as borne out by simulations in the next section, for all cases of interest, *the optically thin and thick regimes correspond directly to weak and strong collisionality respectively*. As a result, we can discuss the frequency scaling of $I_\omega$ in terms of the collisionality.

Figure 1 illustrates the general behavior of the radiation intensity for parameters corresponding to the dielectric breakdown regime of $SiO_2$. For low frequencies ($\omega << \nu$), the radiation intensity increases rapidly with frequency, $I_\omega \sim \omega^2$ (Eq. (11)). The total energy emitted in this range (THz to mid-infrared) is relatively small. Most of the emission originates from the intermediate frequency range ($\omega \sim \nu$), in which $I_\omega$ is nearly constant about its maximum (Eq. (11)). As alluded to in subsection (b) above, this near-constant scaling is limited to a narrow range, $\nu \sim \omega \leq k_B T_e / \hbar$, from the infrared to UV. Finally for large frequencies ($\omega >> \nu$), the exponential factor appearing in Eq. (11) causes a precipitous decrease in $I_\omega$. Overall, the emission spectrum exhibits the following scaling:

$$I_\omega \sim \begin{cases} \omega^2 & \omega << \nu \\ \text{const.} & \omega \sim \nu \\ e^{-\hbar\omega/k_B T_e} & \omega >> \nu \end{cases} . \qquad (13)$$

The absorption length, and hence the optical depth, has a constant scaling for sub-infrared frequencies and falls off rapidly beyond the UV:

$$a_\omega \sim \begin{cases} \text{const.} & \omega << \nu \\ 1/\omega^2 & \omega >> \nu \end{cases} . \qquad (14)$$

The frequency scaling of $I_\omega$ and $a_\omega$ can be understood physically by considering the motion of an electron undergoing collisions in the presence of the radiation field. In the absence of collisions, an electron would exchange no net energy with the radiation field. The electron would oscillate in phase with field, gaining energy in one half-cycle and radiating it away in the next. Collisions break this symmetry: the scattering of the electron from the phonon (or ion) introduces an additional acceleration, such that the electron's oscillation is no longer in phase with the radiation field. This allows net energy exchange. In particular, for fixed radiation field amplitude, the electron excursion in the field, $\delta z_\omega$, decreases with increasing frequency, $\delta z_\omega \sim \omega^{-2}$. In a large frequency field, the small electron excursion makes scattering events unlikely, limiting both absorption and emission. In a low frequency field, on the other hand, the large electron excursion can lead to several collisions before the electron can gain significant energy or coherently emit radiation. As displayed in Fig. (1), the optimum occurs in between. Note that these dynamics are captured implicitly in Eq. (1) and need not be calculated explicitly as with the electron dynamics in the field of the laser pulse.

## 4. Implementation

In this section, we provide working formulas suitable for computation of radiation quantities in a 1D simulation. While the theory above can be implemented in a 3D simulation, the 1D geometry greatly simplifies the often difficult and expensive task of solving a multi-dimensional radiation transport equation, and provides insightful results in a relatively short amount of time. The radiation quantities of interest are based on integrals over the emission coefficient $j_\omega$, which was derived only in the limiting cases of weak and strong collisionality. To provide a smooth transition between these limits, we adopt the formula

$$j_\omega \cong \frac{\alpha}{2\pi^2}\frac{k_B T_e n_e}{m_e c^2}\hbar\nu(T_e)\frac{\omega^2}{\omega^2+\nu^2(T_e)}e^{-\hbar\omega/k_B T_e}. \tag{15}$$

Equation (15) can be readily evaluated given $n_e$ and $T_e$ as a function of space and time.

The monochromatic radiation intensity $I_\omega(t)$ is calculated by solving Eq. (1) in the special case of a semi-infinite slab in the laser pulse propagation direction. In general, the solution of Eq. (1) requires integrating over all paths originating within the dielectric and ending at a particular viewpoint—in our case any point on the dielectric surface (Fig. 2a). However, Apruzese *et al.* have demonstrated that accurate results can be obtained by considering a single representative path at an angle $\theta = \pi/3$ with respect to the axis (Fig. 2b) [24,25]. If $j_\omega$ and $a_\omega$ are known along the path, the solution of Eq. (1) is then straightforward:

$$I_\omega(s_0) = I_\omega(0)e^{-\tau_\omega(0)} + \int_0^{s_0} j_\omega(s)e^{-\tau_\omega(s)}ds, \tag{16}$$

where $I_\omega(s_0)$ is the radiation intensity emerging from the dielectric at the viewpoint $s_0$, $I_\omega(0)$ is the radiation intensity incident on the dielectric from the right, $ds = dx/\cos\theta$ is the optical path length, and $\tau_\omega(s) = \int_s^{s_0} a_\omega(s')ds'$ is the optical depth for the path connecting $s$ and $s_0$. The first term on the right-hand side of Eq. (16) represents attenuation of the incident radiation. For the situation of interest, no such radiation exists and this term is zero. The second term describes the cumulative emission and absorption of bremsstrahlung radiation originating within the dielectric. Equation (16) can be integrated for any time within the simulation. For clarity in the remainder, the monochromatic radiation intensity leaving the dielectric, $I_\omega(s_0)$, will be denoted $I_\omega(t)$.

The integral of $I_\omega(t)$ over time and solid angle yields the spectral fluence (energy emitted per unit area, unit frequency)

$$F_\omega = 2\pi\int_0^{t_{sims}} I_\omega(t)\,dt, \tag{17}$$

while the integral over frequency and solid angle yields the radiation intensity (power emitted per unit area)

$$I(t) = 2\pi\int_0^{\omega} I_\omega(t)\,d\omega. \tag{18}$$

Here the integration over solid angle has been performed over the half-plane to denote radiation escaping the dielectric surface. Because $I_\omega(t)$, $F_\omega$, and $I(t)$ result from spatially integrating $j_\omega(x,t)$, information about the spatial distribution of emitted radiation is lost. We therefore introduce another parameter, the radiation energy density (power emitted per unit volume),

$$E(x) = 2\pi \int_0^{t_{sims}} \int_0^{\infty} j_\omega(x,t)\, d\omega\, dt\,, \tag{19}$$

in order to illustrate the spatial structure of the radiation. For a rough measure of the CBE absorption properties, we also introduce the cumulative, collisional optical depth:

$$\tau_C(t) = \int_0^L a_\omega(x,t)\, dx\ , \tag{20}$$

where $a_\omega(x,t)$ is the absorption coefficient in the collisional regime, Eq. (12). Use of the collisional optical depth is motivated by the preponderance of absorption in the optically thick, collisional regime. This has the added advantage that Eq. (20) is frequency independent, making the optical depth a function of time only. The frequency dependence can, however, be roughly accounted for by setting $\tau_\omega(t) \cong \tau_C(t)/[1+(\omega/\nu)^2]$, providing a reasonable approximation across all collisionality regimes.

Finally, perhaps the most useful, and simplest, quantity for characterizing the radiation emitted from the dielectric is the total radiation energy per unit area: $F_{brems} = \int_0^{t_{sims}} I(t)dt$. With $F_{brems}$ one can calculate the conversion efficiency of laser pulse energy into bremsstrahlung radiation

$$\eta_{brems} = F_{brems} / F_L\,, \tag{21}$$

where $F_L = I_L T_{FWHM}$ is the pulse fluence. Equations (15-21) are used in the numerical simulations described next.

## 5. Simulation results

We now present simulation results of the bremsstrahlung emission from short pulse irradiated $SiO_2$. While the range of pulse intensities and durations of interest can be rather large (several orders of magnitude each), three representative cases are sufficient to illustrate the general behavior. The cases are shown in Fig. 3 in $I_L - T_{\mathrm{FWHM}}$ space. The solid line marks the damage threshold of the dielectric, defined here as the point at which the CBE electron density reaches the critical electron density, i.e. $n_e \cong n_{cr}$ ($n_{cr} = 1.7\times10^{27}\ m^{-3}$ for $\lambda_0 = 800\,\mathrm{nm}$). As one may have expected, the threshold drops with either increasing pulse intensity or duration. Below the damage threshold, the laser pulse photo-excites only a small population of electrons to the conduction band. These electrons, in turn, absorb a small amount of laser energy. Above the damage threshold, the additional photo-excitation and laser pulse heating is sufficient to spark avalanche collisional excitation, which greatly increases the absorbed pulse energy. The distinctness of these two interaction regimes motivates our case selection. Specifically, we chose pulse intensities below (case **1**) and above (case **2**) the damage threshold for a 60 fs FWHM short pulse. A third case (case **3**) was also simulated to compare the bremsstrahlung radiation resulting from a short (60 fs) and long (1 ps) pulse at the same intensity.

The numerical model of the laser pulse-dielectric interaction is the same as in Ref. [22] and similar to that in Ref. [23]. Here we briefly recount the salient features. The model captures the self-consistent propagation of a laser pulse through $SiO_2$ and its interaction with CBEs. The model is 1D+t, resolving dynamics along the laser propagation direction. Three processes of CBE creation and destruction are included: optical field excitation (multiphoton or tunneling), impact excitation, and recombination. The laser pulse can energize the CBEs through inverse bremsstrahlung heating, predominately a result of electron-acoustic phonon scattering. At the same time, this heating depletes the laser pulse energy (similarly with optical field excitation). The CBEs can cool through collisional ionization, recombination, and longitudinal optical phonon excitation. Generally, the CBE cooling through lattice collisions was found to be small; the lattice temperature, as a result, is kept at room temperature. Phase transitions are not accounted for.

The y-polarized laser pulse starts in vacuum and propagates along the x-axis towards a semi-infinite slab of $SiO_2$ occupying the space $x \geq 0$. The newly incorporated formulas to analyze the bremsstrahlung radiation (listed in Section 4) are calculated over the first 500 mm. We have verified that this accounts for most of the radiation.

Figure 4 displays the laser pulse and CBE quantities for case **1** (60 fs, below the damage threshold) at three different times, 0.4, 1.6, and 3.2 ps. The enveloped pulse intensity is shown on the top row. Since the pulse intensity, $I_L$, is below the damage threshold, the pulse traverses the dielectric almost unperturbed (see also Fig. 6 in Ref. [22]). The slight decrease in intensity results from the creation of CBEs through optical field excitation. These electrons have a temperature and density shown in the second and third rows, respectively. The electron temperature peaks at ~4 eV, about half the excitation threshold, 9 eV, and extends over nearly the same region as the laser pulse. The maximum density is rather low, $n_e \cong n_{cr}/100$, and coincides with the location of the pulse. Behind the pulse, the absence of field ionization and heating causes the density to quickly decay. Recombination occurs on a time scale of 150 fs, with a corresponding length of 45 μm, such that the CBE excitation behind the pulse is limited to a few hundred microns.

Figure 5a shows the resulting radiation intensity, and for comparison the laser pulse intensity,

as a function of time. Because the CBEs accumulate over the entire duration of the pulse, the peak of the radiation intensity occurs slightly behind the peak pulse intensity (about 100 fs). The radiation intensity decays over approximately the recombination time as CBEs created at the surface recombine. The radiation persists, however, at a low-level for an interval much longer than the pulse duration, several picoseconds compared to 60 fs. The prolonged low-level emission results from CBEs continually excited as the laser pulse traverses the dielectric. The low-level emission eventually decays as photo-excitation depletes the laser pulse energy. In particular, a small reduction in pulse intensity causes a large drop in the excitation rate (approximately $I_L^6$), and since $j_\omega \sim n_e$, a drop in radiation intensity follows.

The spatial distribution of the radiation energy density created as the laser pulse traverses the dielectric can be observed in Fig. 5b. Perhaps the most important consequence of the processes described above is *volumetric* emission from the dielectric, i.e. a significant portion of the radiation originates from *within* the dielectric, not the surface. The energy density decreases exponentially away from the surface, albeit slowly, over ~100 mm. About half of the radiation is emitted from the region $x = 0$ to $x = 50\ \mu\text{m}$, and half from $x > 50\ \mu\text{m}$.

The CBEs can absorb some of the radiation they emit. The collisional optical depth (Eq. (20)) is shown in Fig. 5c. The moderate optical depth, ~4, early in the emission demonstrates that absorption does take place, but is not significant. We note that Eq. (20) accounts for absorption by the CBEs only, and that the dielectric itself may also absorb radiation; this effect is not included in our model.

The time-integrated radiation spectrum emerging from the dielectric (the spectral fluence $F_\omega$, Eq. (17)) is plotted in Fig. 5d. The spectrum resembles that in Fig. 1. Even though Fig. 5d was generated with parameters below the damage threshold and Fig. 1 with parameters above, the two exhibit the same frequency scaling. The spectrum peaks at a frequency about twice the laser frequency ($\lambda_{\text{peak}} = 450\text{ nm}$), the latter shown with a dashed vertical line. In the ultraviolet, the intensity falls sharply on account of the $\exp(-\hbar\omega / k_B T_e)$ factor: due to the low temperature, few electrons have sufficient energy to emit a UV photon.

To further illustrate the extent of absorption, the spectral fluence calculated without radiation transport is also plotted (dashed curve). In the strongly collisional regime ($\omega << \nu$), absorption by the CBEs attenuates the spectral fluence by a factor corresponding roughly to the optical depth in Fig. 5c. This is in qualitative agreement with Eq. (8), which is valid only for a uniform CBE density. In the opposite regime of weak collisionality ($\omega >> \nu$), the attenuation is negligible. Here the CBEs appear optically thin to the high-frequency radiation, $\tau_\omega \sim \omega^{-2} << 1$, and the spectrum mirrors Eq. (7).

When the laser pulse fluence exceeds the damage threshold, it significantly modifies the laser pulse-dielectric interaction. As shown in Fig. 6, which displays the results for case **2** (60 fs, above threshold), enhanced energy absorption by the CBEs exhausts the pulse energy at a much faster rate. Halfway through the dielectric, the pulse intensity has dropped by an order of magnitude, from $2\times10^{18}$ to $2.5\times10^{17}$ W/m$^2$ (Fig. 6b). The electron temperature, however, increases only marginally, from about 4 to 5 eV (c.f. Figs. 4d-f & 6d-f). The extra deposited laser pulse energy goes into increasing the electron density. As a result, a thin sheath of CBEs with density exceeding the critical density forms on the dielectric surface (Fig. 6g), and gradually decays through recombination (Fig. 6h). In addition, the transmitted, now low intensity pulse excites a small number of CBEs inside the dielectric, akin to case **1**.

Analogous to Fig. 5, Fig. 7 displays the bremsstrahlung radiation quantities for case **2**. Several differences in the below and above threshold radiation quantities are worth noting. First, the radiation intensity plotted in Fig. 7a has similar time dependence to that in Fig. 5a, but its magnitude is substantially larger, about three orders of magnitude. The radiation energy density (Fig. 7b), on the other hand, has an entirely different structure. The emission originates almost entirely from a thin layer (<<1 mm) on the dielectric surface. That is above the damage threshold the dielectric is a *surface* emitter, while below it is a *volumetric* emitter. Other than the ~$10^3$ increase in magnitude, the spectrum (Fig. 7d) resembles that of case **1**. Again the CBEs appear optically thick and thin in the strongly and weakly collisional regimes respectively.

Figures 8 and 9 show the simulation results for case **3** (1 ps, above threshold). Early in the pulse, the CBE density at the dielectric surface reaches the critical density, $n_e > n_{cr}$. As a result, the surface reflects most of the laser pulse energy. Otherwise the features of the CBEs are similar to case **2**: a thin CBE surface layer and a rapidly decaying CBE density behind the pulse. Unlike runs **1** and **2** however, significant bremsstrahlung emission occurs over a picosecond, an interval comparable to the pulse duration (Fig. 9a). In general, the greater of either the recombination time or pulse duration limits the interval over which the CBEs emit significant bremsstrahlung radiation.

The radiation energy density has the same spatial structure as case **2** (c.f. Figs. 7b & 9b). For both cases, $I_L >> I_{thr}$, and the bremsstrahlung radiation originates from the dielectric surface. A comparison of Figs. 7d & 9d demonstrates that the pulse duration has little impact on the radiation spectrum. This can be explained by the nearly constant peak electron temperature in both cases (3-5 eV), which determines the spectral peak of the radiation.

The total conversion efficiency of laser pulse energy into bremsstrahlung radiation as a function of peak intensity is plotted in Fig. 10 for both a short (60 fs) and long (1 ps) pulse. The conversion efficiency increases exponentially with $I_L$, saturates, then increases sharply near the damage threshold, until it saturates again. The exponential rise for $I_L < I_{thr}$ results from the nonlinear increase in electron density with laser intensity: $F_{brems} \sim n_e \sim I_L^6$. The first saturation occurs as the laser intensity approaches $I_{thr}$, and results from the enhanced reflection of the laser pulse as the surface CBE density increases. For $I_L > I_{thr}$, avalanche impact excitation occurs causing an enormous growth of the CBE density at the surface. Finally, when all the electrons are excited from valence, the electron density in the conduction band saturates and the conversion efficiency levels off. Both the short and long pulses exhibit similar behavior.

## 6. Conclusions

We have developed a theory of bremsstrahlung emission appropriate for ultrashort laser pulse-dielectric interactions. The theory is valid for arbitrary ratio of collision-to-radiation frequency, $\nu / \omega$, and can, in principle, be generalized to any type of electron collision. Salient features of the radiation spectrum were established, including frequency and temperature dependence. The theory was integrated into a laser pulse-dielectric interaction simulation, which provided examples of the expected emission spectra, power density, energy, and conversion efficiency. The most important findings of this paper are:

- The emission may originate either from the "surface" or the bulk of the dielectric depending on the laser intensity with respect to damage threshold.
- The bremsstrahlung radiation spectrum increases as $\omega^2$, peaks near the visible and drops sharply for frequencies $\hbar\omega >> k_B T_e$. The general shape of the spectrum is not sensitive to the laser pulse parameters (intensity and pulse duration).
- The conversion efficiency of laser pulse energy into bremsstrahlung radiation is a complicated function of the pulse intensity. It increases sharply with $I_L$, and saturates twice: once near $I_{thr}$ due to reflection from CBEs generated at the dielectric surface, and again when all the surface electrons are transferred from valence to the conduction band. The maximum conversion efficiency calculated in this work is about $\sim 10^{-5}$.
- The optically thin and thick regimes correspond directly to the weakly and strongly collisional

regimes respectively. The transition occurs near the optical frequency range.

**Acknowledgements**

The authors would like to thank J. Giuliani, B. Rock, M. Helle, and L. Johnson for fruitful discussions. This work was supported by the US Naval Research Laboratory 6.1 Base Program and the Office of Naval Research.

**Figure captions:**

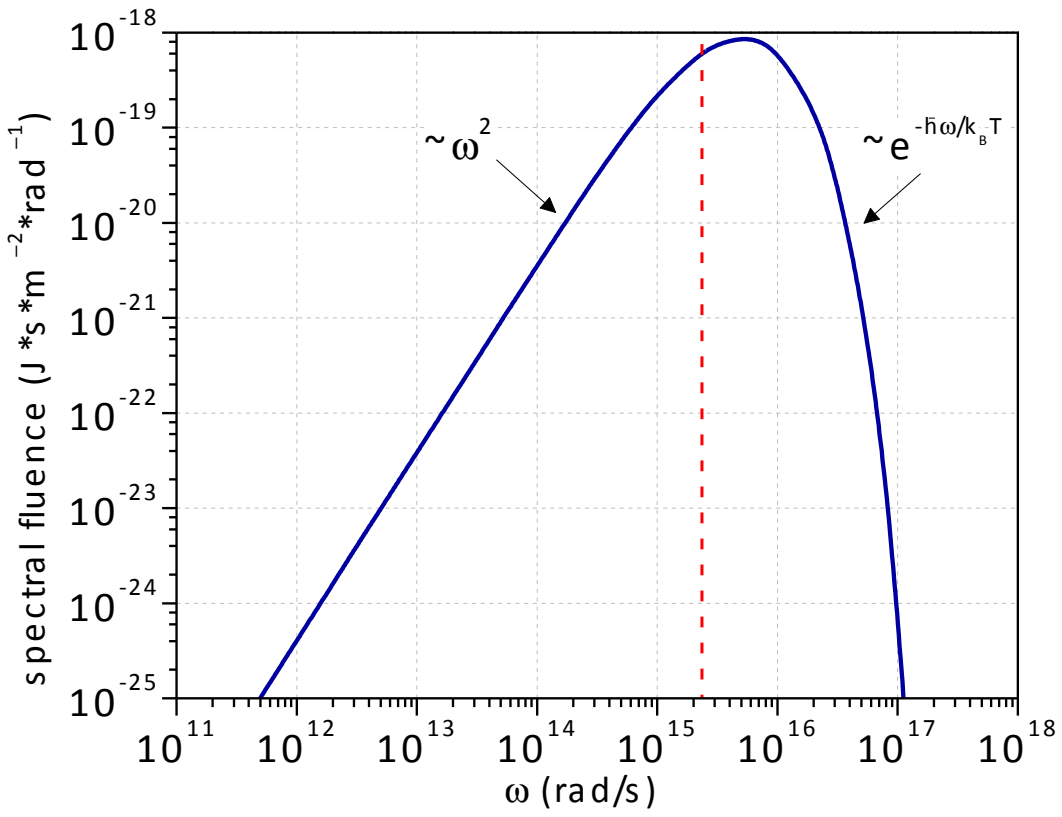

Figure 1. Spectral fluence of the bremsstrahlung radiation created during the interaction of a laser pulse with $SiO_2$ near breakdown. The pulse intensity, duration, and wavelength are $I_L = 6\times10^{17}\ \mathrm{W\cdot cm^{-2}}$, $T_{FWHM} = 60\ \mathrm{fs}$, and $\lambda_L = 800\ \mathrm{nm}$, respectively. The dashed vertical line corresponds to the laser frequency.

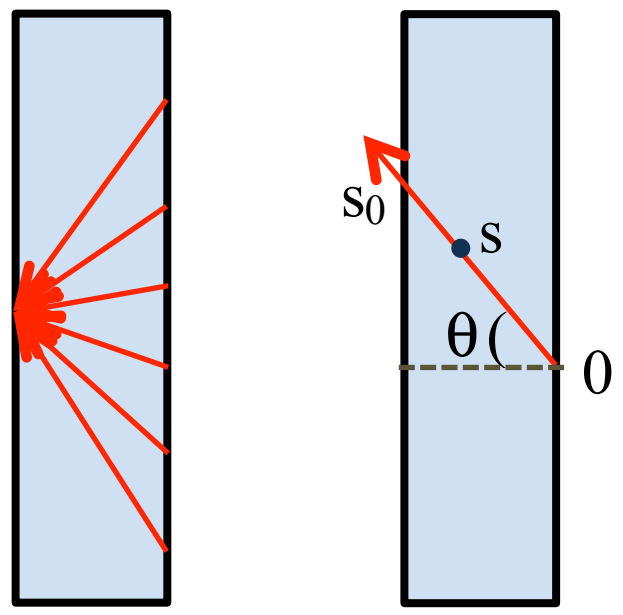

Figure 2. Schematics of radiation transport calculations. (a) conventional radiation transport treatment with a set of paths at various angles; (b) representative path used in the simulations.

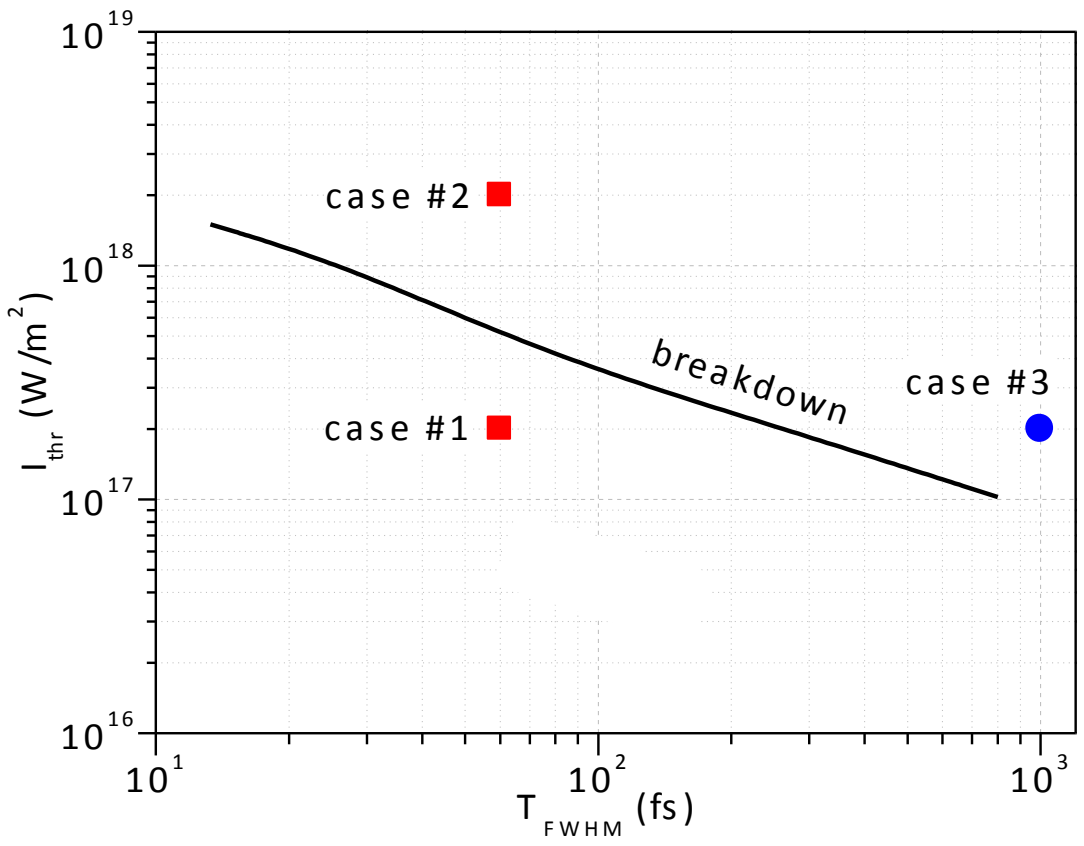

Figure 3. Threshold intensity for versus laser pulse duration (solid line). Symbols denote the locations of the runs. Case **1**: $I_L = 2\times10^{17}\ \mathrm{W\cdot cm^{-2}}$, $T_{FWHM} = 60$ fs, Case **2**: $I_L = 2\times10^{18}\ \mathrm{W\cdot cm^{-2}}$, $T_{FWHM} = 60$ fs, Case **3**: $I_L = 2\times10^{17}\ \mathrm{W\cdot cm^{-2}}$, $T_{FWHM} = 1$ ps.

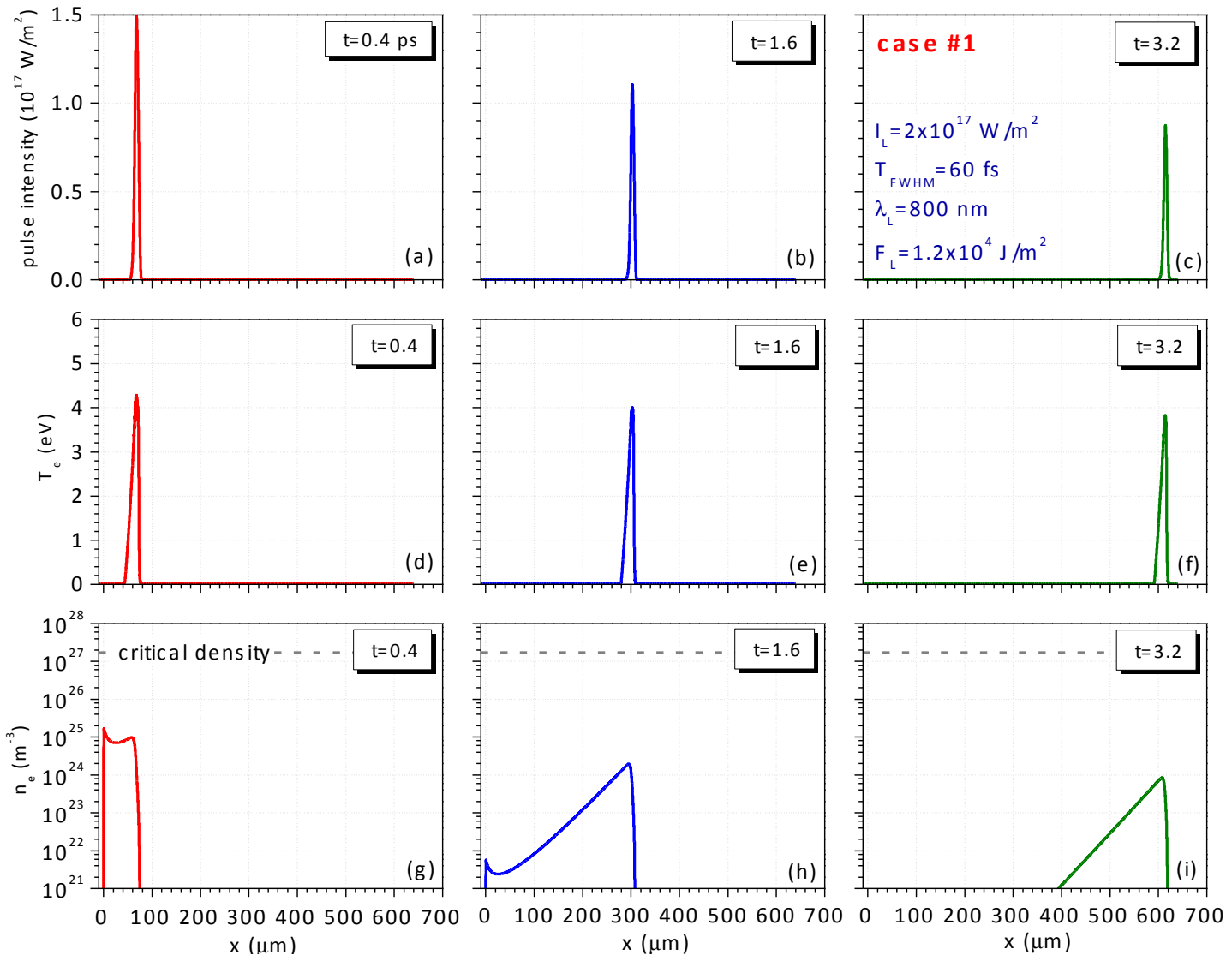

Figure 4. Calculated pulse intensity (top row), electron temperature (middle row) and electron density (bottom row) for case **1**.

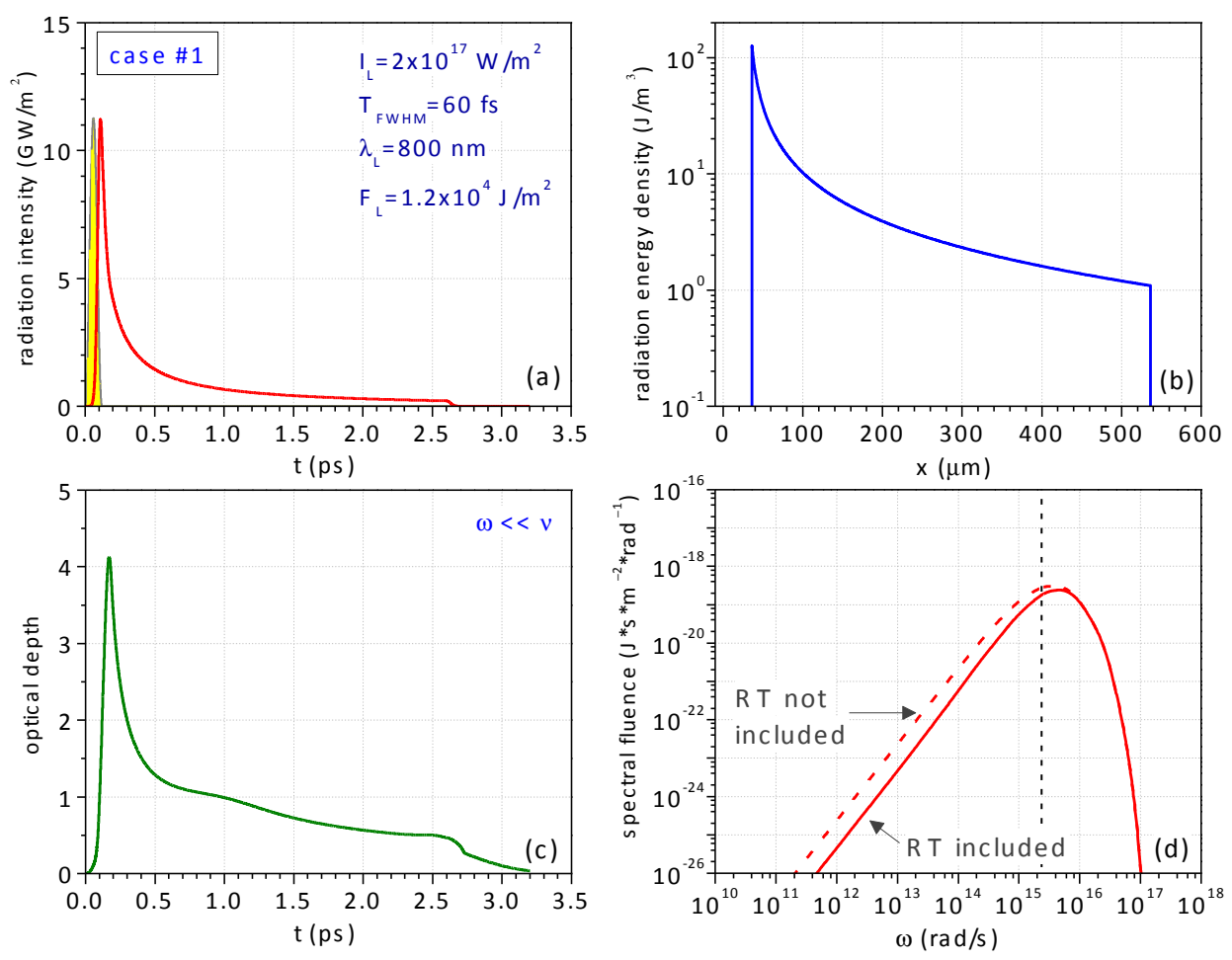

Figure 5 Radiation intensity (a), radiation energy density (b), optical depth (c), and spectral fluence (d) for case **1**. The dielectric length is 500 mm.

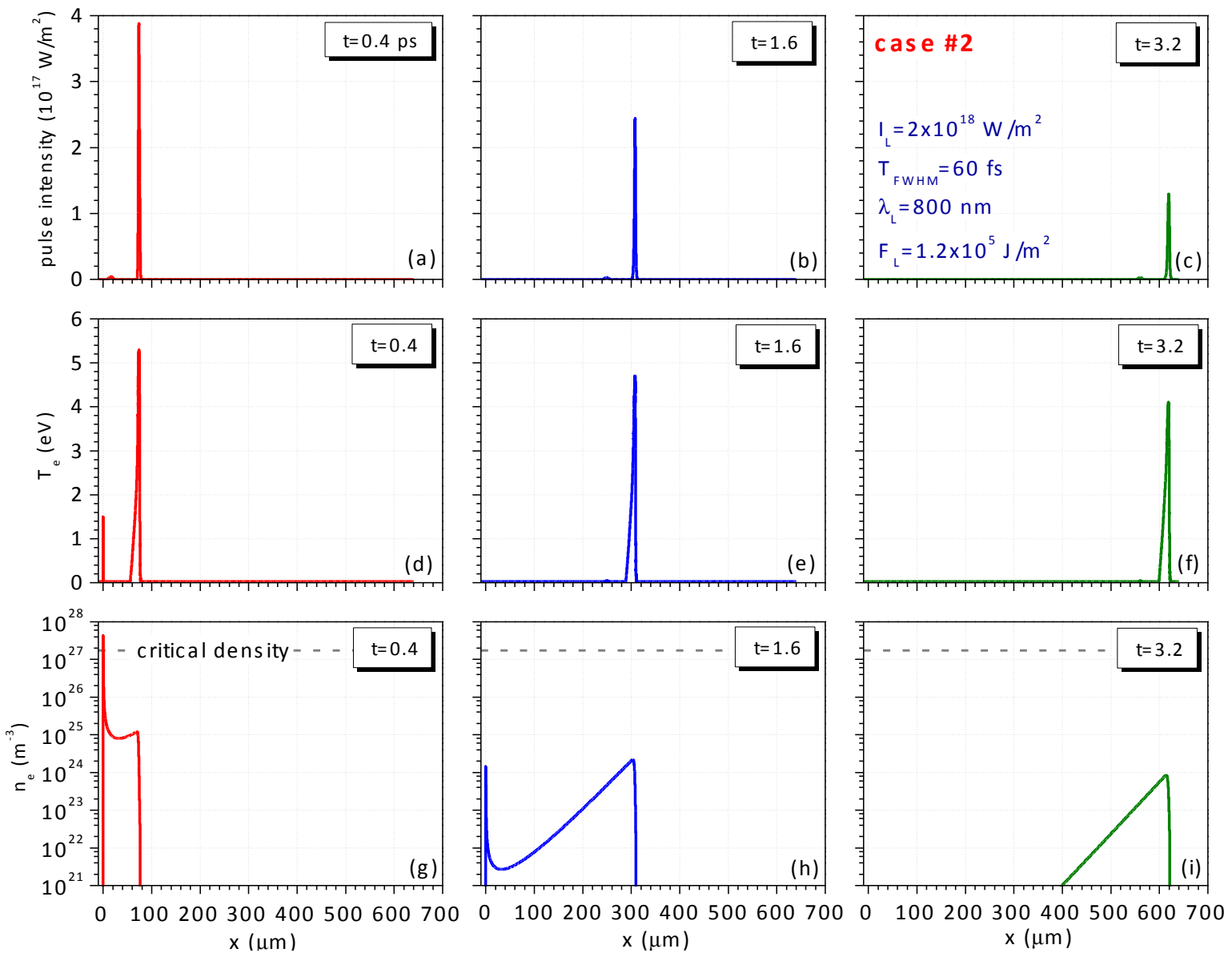

Figure 6. Calculated pulse intensity (top row), electron temperature (middle row) and electron density (bottom row) for case **2**.

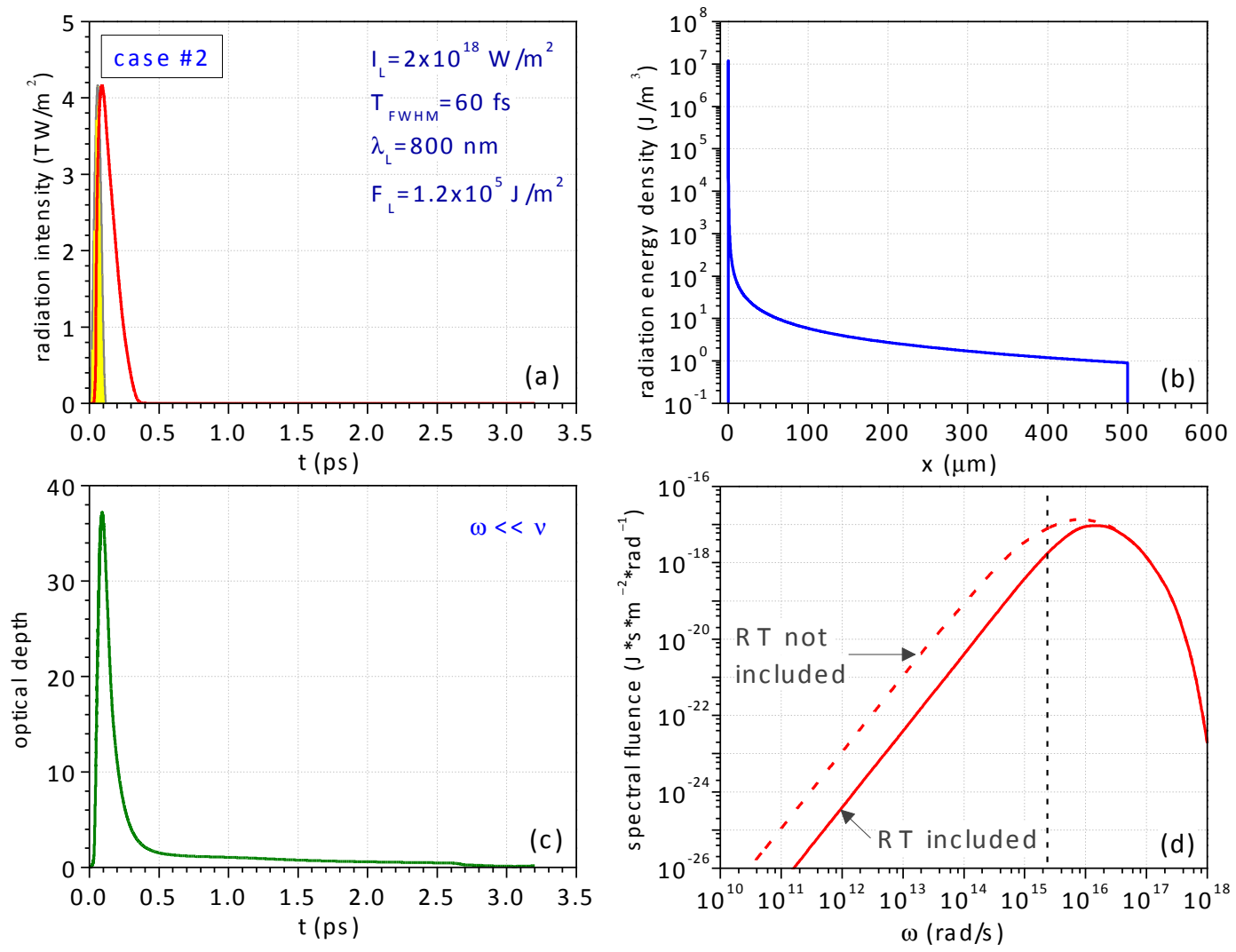

Figure 7 Radiation intensity (a), radiation energy density (b), optical depth (c), and spectral fluence (d) for case **2**. The dielectric length is 500 mm.

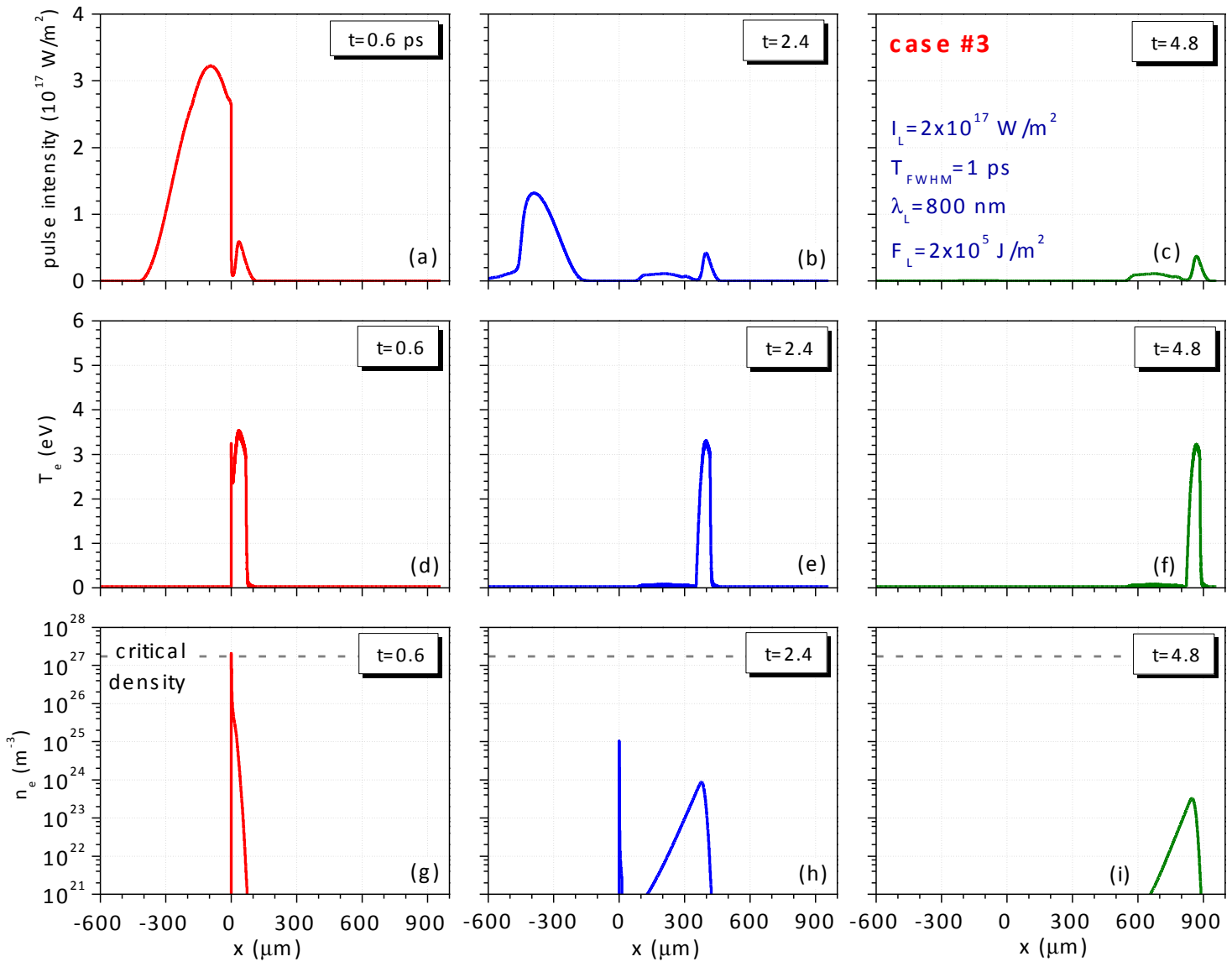

Figure 8. Calculated pulse intensity (top row), electron temperature (middle row) and electron density (bottom row) for case **3**.

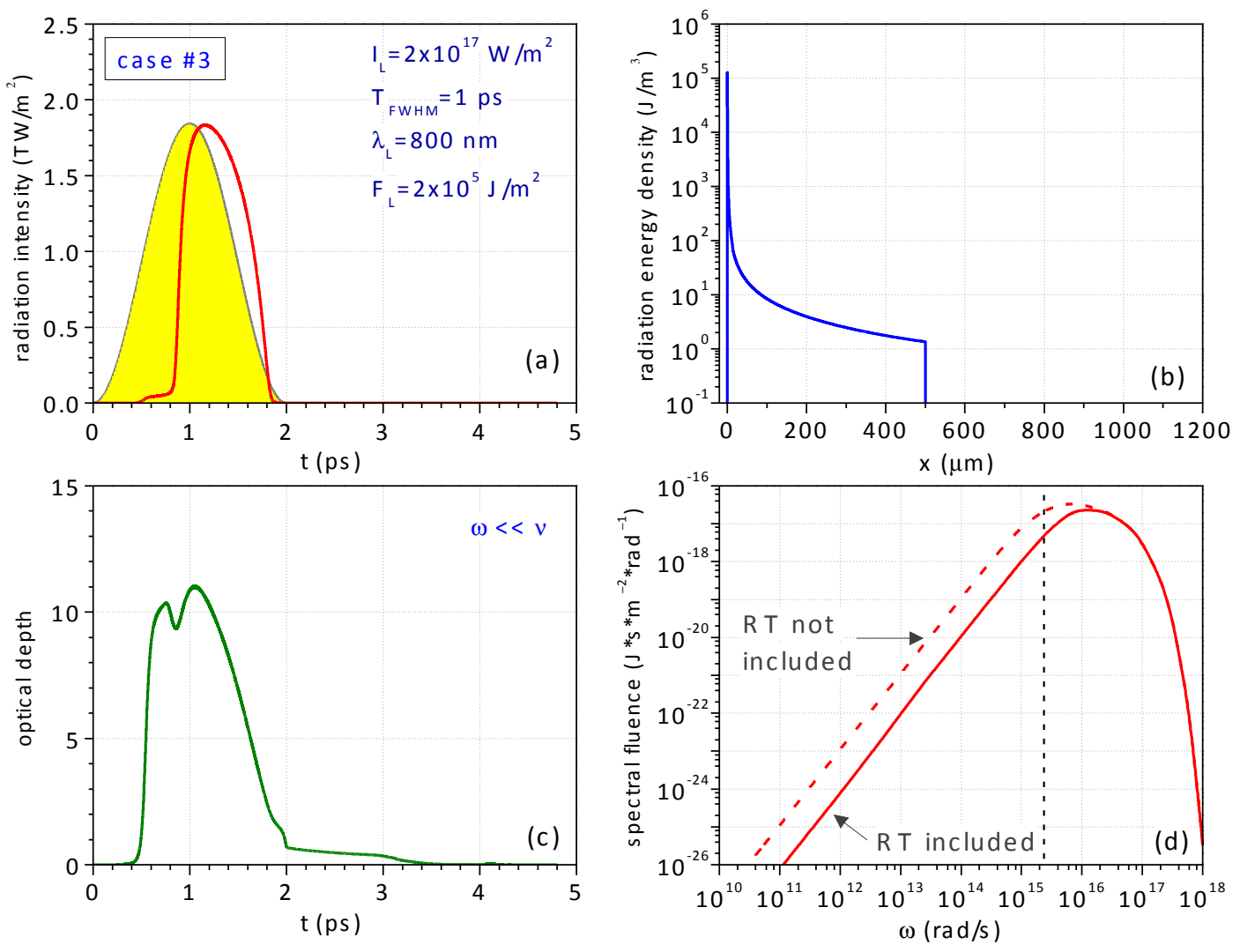

Figure 9. Radiation intensity (a), radiation energy density (b), optical depth (c), and spectral fluence (d) for case **3**. The dielectric length is 500 mm.

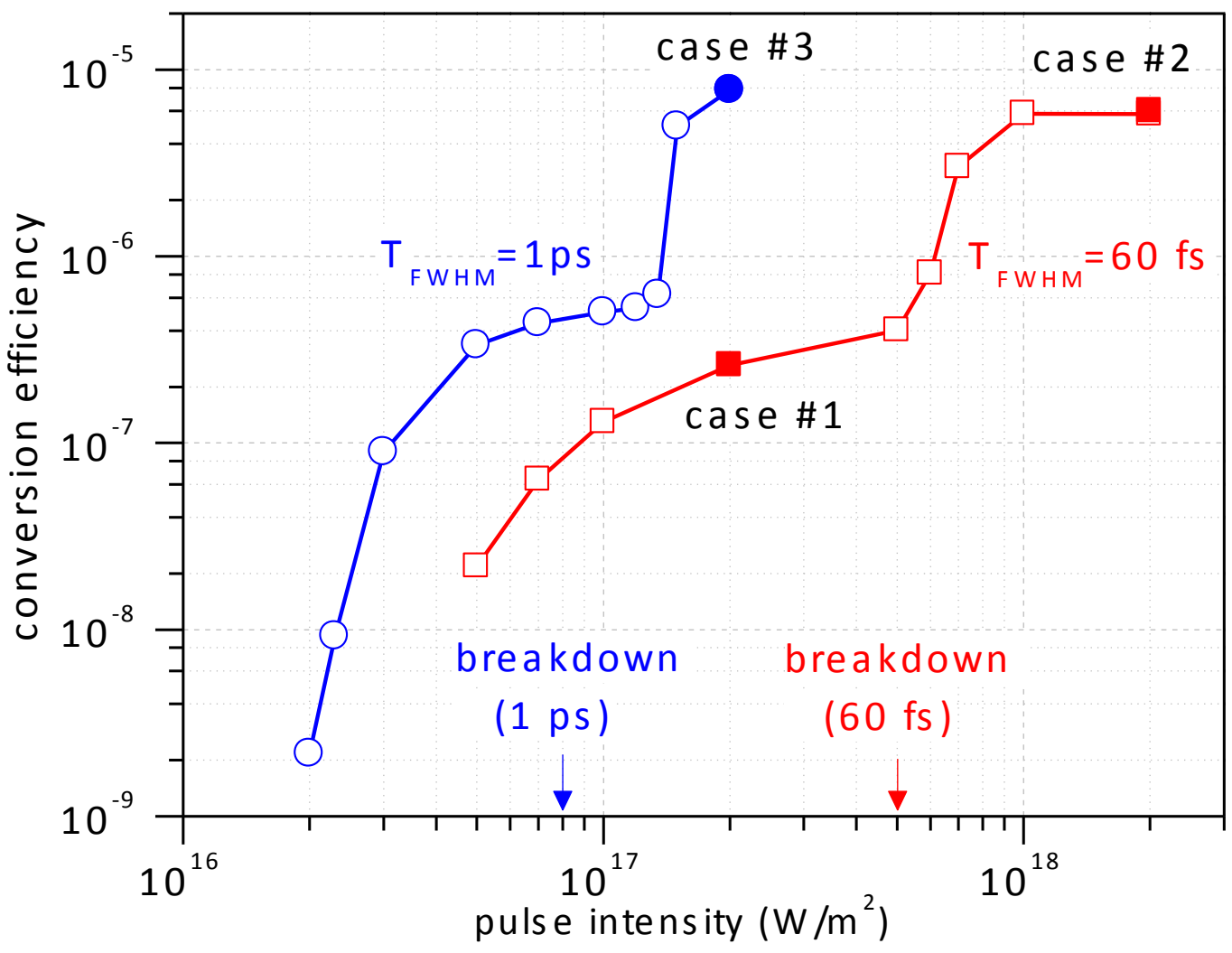

Figure 10. Conversion efficiency of laser pulse energy into bremsstrahlung radiation versus $I_L$ for durations $T_{FWHM} = 60$ fs and $T_{FWHM} = 1$ ps . The arrows denote the breakdown threshold.

Table 1. A summary of bremsstrahlung radiation formulas in the limiting cases of weak and strong collisionality.

| **parameter** | **units** | **n <<w** | **n>>w** |
|---|---|---|---|
| $\eta_\omega(\varepsilon,\omega)$ | J/(sr*rad) | $\frac{\alpha}{3\pi^2}\frac{\varepsilon}{m_e c^2}\mathrm{h}\nu(\varepsilon)$ | $\frac{\alpha}{3\pi^2}\frac{\varepsilon}{m_e c^2}\frac{\mathrm{h}\omega^2}{\nu(\varepsilon)}$ |
| $j_\omega(\omega)$ | J/(m$^3$*sr*rad) | $\frac{\alpha}{2\pi^2}\frac{k_B T_e n_e}{m_e c^2}\mathrm{h}\nu(T_e)e^{-\mathrm{h}\omega/k_B T_e}$ | $\frac{\alpha}{2\pi^2}\frac{k_B T_e n_e}{m_e c^2}\frac{\mathrm{h}\omega^2}{\nu(T_e)}e^{-\mathrm{h}\omega/k_B T_e}$ |
| $a_\omega(\omega)$ | 1/m | $\frac{\omega_p^2}{\omega^2}\frac{\nu(T_e)}{c}$ | $\frac{\omega_p^2}{c\nu(T_e)}$ |
| $\boldsymbol{j}$ | W/(m$^3$*sr) | $\frac{\alpha}{2\pi^2 m_e c^2}\left(k_B T_e\right)^2\nu(T_e)n_e$ | $\frac{\alpha}{\pi^2\mathrm{h}^2 m_e c^2}\frac{\left(k_B T_e\right)^4}{\nu(T_e)}n_e$ |